\documentclass[conference]{IEEEtran}
\usepackage{amssymb}
\usepackage{enumerate}
\usepackage{graphicx}
\usepackage{multirow}
\usepackage{arydshln}
\usepackage{color}
\usepackage{float}
\usepackage{bm}                                                                   
\usepackage{comment}
\usepackage{cite}
\usepackage{amsmath}
\usepackage{algorithmic}
\usepackage{array}

\newcommand{\argmax}{\mathop{\rm arg~max}\limits} 

\newlength\savedwidth
\newcommand{\wcline}[1]{\noalign{\global\savedwidth\arrayrulewidth\global\arrayrulewidth 1.0pt} \cline{#1}
\noalign{\global\arrayrulewidth\savedwidth}}

\ifCLASSINFOpdf
  \usepackage[pdftex]{graphicx}
\else
\fi
\begin{document}
%
\title{Acoustic Scene Analysis Using Partially Connected\\Microphones Based on Graph Cepstrum}

\author{\IEEEauthorblockN{Keisuke Imoto}
\IEEEauthorblockA{Ritsumeikan University, Japan}}

\maketitle
\begin{abstract}
In this paper, we propose an effective and robust method for acoustic scene analysis based on spatial information extracted from partially synchronized and/or closely located distributed microphones.
In the proposed method, to extract spatial information from distributed microphones while taking into account whether any pairs of microphones are synchronized and/or closely located, we derive a new cepstrum feature utilizing a graph-based basis transformation.
Specifically, in the proposed graph-based cepstrum, the logarithm of the amplitude in a multichannel observation is converted to a feature vector by an inverse graph Fourier transform, which can consider whether any pair of microphones is connected.
Our experimental results indicate that the proposed graph-based cepstrum effectively extracts spatial information with consideration of the microphone connections.
Moreover, the results show that the proposed method more robustly classifies acoustic scenes than conventional spatial features when the observed sounds have a large synchronization mismatch between partially synchronized microphone groups.
%
\end{abstract}
\IEEEpeerreviewmaketitle
\section{Introduction}
\label{sec:intro}
Acoustic scene analysis (ASA), which analyzes scenes in which sounds are produced, is now a very active research area in acoustics, and it is expected that ASA will enable many useful applications such as systems monitoring elderly people or infants \cite{Peng_ICME2009_01,Guyot_ICASSP2013_01}, automatic surveillance systems \cite{Harma_ICME2005_01,Radhakrishnan_WASPAA2005_01,Ntalampiras_ICASSP2009_01,Komatsu_ICASSP2017_01}, automatic file-logging systems \cite{Eronen_TASLP2006_01,Imoto_IEICE2016_01,Schroder_ICASSP2016_01}, and advanced multimedia retrieval \cite{Zhang_TASLP2001_01,Jin_INTERSPEECH2012_01,Ohishi_ICASSP2013_01,Liang_ICASSP2017_01}.

To analyze scenes from an acoustic signal, many approaches based on machine learning techniques have been proposed.
For instance, Eronen {\it et al.} \cite{Eronen_TASLP2006_01} and Mesaros {\it et al.} \cite{Mesaros_EUSIPCO2010_01} have proposed spectral feature-based methods such as mel-frequency cepstral coefficients (MFCCs) and Gaussian mixture models (GMMs).
Han {\it et al.} \cite{Han_DCASE2017_01} and Jallet {\it et al.} \cite{Jallet_DCASE2017_01} have proposed methods using the mel-spectrogram as input features and the convolutional neural network (CNN) or recurrent convolutional neural network (RCNN) as classifiers.
Guo and Li \cite{Guo_JNN2003_01}, Kim {\it et al.} \cite{Samuel_WASPAA2009_01}, and Imoto and co-workers \cite{Imoto_IEICE2016_01,Imoto_MLSP2013_01} have investigated ASA utilizing intermediate feature representations based on acoustic event histograms.

ASA based on spatial information extracted from a microphone array composed of smartphones, smart speakers, and IoT devices has also been proposed \cite{Kwon_ISCS2009_01,Phan_WASPAA2015_01,Giannoulis_EUSIPCO2015_01}.
Many of these methods extract spatial information based on observed time differences or sound power ratios between channels, and therefore, they require that the microphones are synchronized and the microphone locations and array geometry are known.
However, since the distributed microphones in multiple smartphones, smart speakers, or IoT devices are often unsynchronized and the microphone locations and array geometry are unknown, conventional methods cannot be applied to such distributed microphone arrays.
To extract spatial information using unsynchronized distributed microphones whose locations and array geometry are unknown, Imoto and Ono have proposed a spatial cepstrum that can be applied under these conditions \cite{Imoto_TASLP2017_01}.
In this approach, log-amplitudes obtained by multiple microphones are converted to a feature vector similarly to when using the cepstrum, which is based on principal component analysis (PCA) of the feature vector.

\begin{figure}[t]
\begin{center}
\includegraphics[width=0.95\columnwidth]{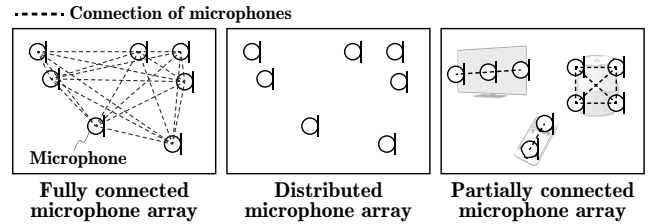}
\vspace{-10pt}
\end{center}
\caption{Example of microphone connections}
\label{fig:connection01}
\end{figure}

On the other hand, the numbers of smartphones, smart speakers, or IoT devices that have multiple microphones have been increasing.
A microphone array composed of these microphones are often partially synchronized or closely located as shown in Fig.~\ref{fig:connection01}; we refer to these synchronized or closely located microphones collectively as {\it connected} microphones.
The time delay or sound power ratio between channels is a significant cue for extracting spatial information even when the microphones are partially connected; however, the conventional spatial cepstrum does not consider whether some of the microphones are partially connected.

%

In this paper, we propose a novel spatial feature extraction method for a distributed microphone array that can take into account whether or not microphones are partially connected.
To consider whether any pairs of microphones are connected, we utilize a graph representation of the microphone connections, where the power observations and microphone connections are represented by the weights of the nodes and edges, respectively. 
Then, the proposed method introduces a graph Fourier transform, which enables spatial feature extraction considering the connections between microphones.

This paper is organized as follows.
In section 2, the spatial cepstrum used in conventional spatial feature extraction for a distributed microphone array is introduced.
In section 3, the proposed method of extracting a spatial feature for partially connected distributed microphones and the similarity of the proposed method to the conventional cepstrum and spatial cepstrum are discussed.
In section 4, experiments performed to evaluate the proposed method are reported.
In section 5, we conclude this paper.
%
%
\section{Conventional Spatial Feature Extraction for Distributed Microphones}
\label{sec:SC}
To extract spatial information from unsynchronized distributed microphones whose locations and array geometry are unknown, the spatial cepstrum, which is a similar technique to the {\it cepstrum} feature, has been proposed \cite{Imoto_TASLP2017_01}.

Suppose that a multichannel observation is recorded by $N$ microphones and $\bar{a}_{\tau,n}$ denotes the power observed for microphone $n$ at time frame $\tau$.
In the case of unsynchronized distributed microphones, synchronization over channels is still a challenging problem and phase information may be unreliable.
Therefore, the spatial cepstrum utilizes only the log-amplitude vector
%
\begin{align}
  {\bf q}_{\tau} = \left(
  \begin{array}{c}
    \log \bar{a}_{\tau,1} \\
    \log \bar{a}_{\tau,2} \\
    \vdots \\
    \log \bar{a}_{\tau,n} \\
    \vdots \\
    \log \bar{a}_{\tau,N}
  \end{array}
  \right),
\end{align}
%
\noindent which is relatively robust to a synchronization mismatch.
Considering that the distributed microphones may be non-uniformly located, PCA is then applied for the basis transformation of the spatial cepstrum instead of the inverse discrete Fourier transform (IDFT).
Suppose that ${\bf R}_q$ is the covariance matrix of ${\bf q}_\tau$ and given by
%
\begin{align}
  {\bf R}_q = \frac{1}{T}\sum_\tau {\bf q}_\tau{\bf q}_\tau^{\mathsf{T}},
\end{align}
%
\noindent where $T$ is the number of time frames.
Since ${\bf R}_{q}$ is a symmetric matrix, the eigendecomposition of ${\bf R}_{q}$ can be represented as
%
\begin{align}
  {\bf R}_{q} &= {\bf E} {\bf D} {\bf E}^{\mathsf{T}},
\end{align}
%
\noindent where ${\bf E}$ and ${\bf D}$ are the eigenvector matrix and the diagonal matrix whose diagonal elements are equal to the eigenvalues in descending order, respectively.
Using this eigenvector matrix ${\bf E}$, the spatial cepstrum is defined as
%
\begin{align}
  {\bf d}_{\tau}={\bf E}^{\mathsf{T}} {\bf q}_{\tau}.
  \label{eq:def_SC}
\end{align}
%
The spatial cepstrum can extract spatial information without microphone locations or the array geometry, although it requires training sounds to estimate the eigenvector matrix ${\bf E}$ by PCA.
Moreover, since the spatial cepstrum does not consider whether or not the microphones are connected, observed time differences or sound power ratios between channels cannot be utilized for spatial feature extraction.
%
\section{Spatial Feature Extraction Based on Graph Cepstrum}
\label{sec:GFTceps}
\subsection{Graph Cepstrum}
\label{subsec:gc}
We consider the situation that a microphone array is composed of multiple generic acoustic sensors mounted on smartphones, smart speakers, or IoT devices, where some of the microphones mounted on each device are connected.
To extract spatial information while considering microphone connections, we here propose a novel spatial feature extraction method that utilizes a graph representation of the multichannel observations and microphone connections.
Specifically, to extract spatial information, the proposed method performs the graph Fourier transform \cite{Ribeiro_EUSIPCO2016_01} instead of PCA in the spatial cepstrum.
This makes it possible to take into account which pairs of microphones are connected.

%
\begin{figure}[t]
\begin{center}
\includegraphics[width=0.95\columnwidth]{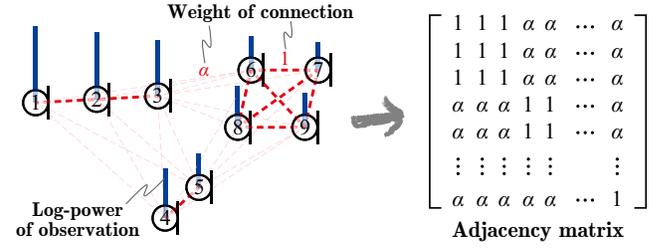}
\vspace{-12pt}
\end{center}
\caption{Example of observations on graph and relationship between microphone connections and adjacency matrix}
\label{fig:adjacency01}
\end{figure}

Consider the logarithm powers of the observations on the graph shown in Fig.~\ref{fig:adjacency01}, where the power observations and microphone connections are represented by the weights of the nodes and edges, respectively. 
Here, the $N \times N$ adjacency matrix is defined as
%
\begin{align}
{\bf A}(m,n) = \begin{cases}
    1 & {\rm if\ channels}\ m\ {\rm and}\ n\ {\rm are\ connected}\\
    0 \ {\rm or} \ \alpha & otherwise, \end{cases}
\end{align}
%
where $\alpha$ is an arbitrary weight of the connection within the range of 0.0--1.0.
We also assume the $N \times N$ degree matrix {\bf D}, which is a diagonal matrix whose diagonal elements are represented as
%
\begin{align}
{\bf D}(m,m) &= \sum_{n} {\bf A}(m,n).
\end{align}
%
\noindent The degree matrix indicates the number of microphones connected with microphone $m$.
Then, the unweighted graph Laplacian is written as
%
\begin{align}
{\bf L} &\triangleq {\bf D} - {\bf A},
\end{align}
%
\noindent where {\bf L} is also a symmetric matrix since both {\bf D} and {\bf A} are symmetric matrices.
Thus, eigendecomposition of ${\bf L}$ can be expressed as
%
\begin{align}
  {\bf L} &= {\bf U} {\bf \Lambda} {\bf U}^{\mathsf{T}},
\end{align}
%
\noindent where ${\bf U}$ and ${\bf \Lambda}$ are the eigenvector matrix and the diagonal matrix whose diagonal elements $\lambda_{m}$ are equal to the eigenvalues in ascending order, respectively.
The eigenvector matrix ${\bf U}^{\mathsf{T}}$ and its transpose ${\bf U}$ are the graph Fourier transform (GFT) matrix and the inverse graph Fourier transform (IGFT) matrix, respectively, which enable the basis transformations considering the connections between microphones.

Thus, the proposed spatial feature, which can consider the connections between microphones, is defined in terms of the IGFT of the log-amplitude vector ${\bf q}_{\tau}$ as

\vspace{-14pt}
\begin{align}
  {\bf e}_{\tau}={\bf U} {\bf q}_{\tau}.
  \label{eq:def_gSC}
\end{align}
\vspace{-14pt}

Because this proposed spatial feature also resembles the conventional cepstrum as well as the spatial cepstrum, we call it the graph cepstrum (GC).
%
%
\begin{figure}[t]
\vspace{2pt}
\begin{center}
\includegraphics[width=0.80\columnwidth]{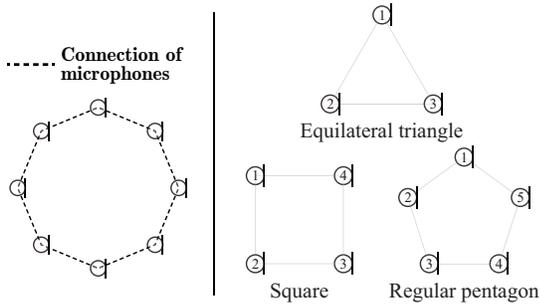}
\vspace{-3pt}
\end{center}
\caption{Examples of ring graph condition (left) and circularly symmetric microphone arrangements (right)}
\label{fig:symmetric01}
\end{figure}
%
%
\subsection{Graph Cepstrum on Ring Graph}
\label{sec:Spcase}
Let us consider a circular connected condition, namely the ring graph condition shown in Fig.~\ref{fig:symmetric01}.
For this condition, a graph Laplacian is represented as the circulant matrix
%
\begin{align}
\hspace*{-10pt} {\bf L}_{sym} = \left[ \begin{array}{rrrcrrr}
2&-1&0&\ \ \cdots\!&0&0&-1\ \ \\
-1&2&-1&\ \ \cdots\!&0&0&0\ \ \\
0&-1&2&\ \ \cdots\!&0&0&0\ \ \\
\vdots&\vdots&\vdots&\ \ \ddots\!&\vdots&\vdots&\vdots\ \ \\
0&0&0&\ \ \cdots\!&2&-1&0\ \ \\
0&0&0&\ \ \cdots\!&-1&2&-1\ \ \\
-1&0&0&\ \ \cdots\!&0&-1&2\ \ \end{array} \right]. \hspace{-6pt}
\label{eq:symL}
\end{align}
%
On the basis of the fact that a circulant matrix is diagonalized by an IDFT matrix ${\bf Z}_{N}$ \cite{Golub_JHUnivPress1996_01} defined by
%
\begin{align}
{\bf Z}_{N} \! &= \! \frac{1}{\sqrt{N}} \! \left[ \begin{array}{cccccc} \!1 \!&\! 1 \!\!&\!\! 1 \!&\!\! \cdots \!\!&\! 1 \!\!&\!\! 1\\
\!1 \!&\! \zeta^{1} \!\!&\!\! \zeta^{2} \!\!&\!\!\! \cdots \!\!\!&\!\! \zeta^{N\!-\!2} \!\!&\!\! \zeta^{N-1}\\
\!1 \!&\! \zeta^{2} \!\!&\!\! \zeta^{4} \!\!&\!\!\! \cdots \!\!\!&\!\! \zeta^{2(N\!-\!2)} \!\!&\!\! \zeta^{2(N\!-\!1)}\\
\!\vdots &\! \vdots \!\!&\!\! \vdots \!\!&\!\!\! \ddots \!\!\!&\!\! \vdots \!\!&\!\! \vdots\\
\!1 \!&\! \zeta^{N\!-\!2} \!\!&\!\! \zeta^{2(N\!-\!2)} \!\!&\!\!\! \cdots \!\!\!&\!\! \zeta^{(N-2)^2} \!\!&\!\! \zeta^{(N\!-\!1)(N\!-\!2)}\\
\!1 \!&\! \zeta^{N\!-\!1} \!\!&\!\! \zeta^{2(N\!-\!1)} \!\!&\!\!\! \cdots \!\!\!&\!\! \zeta^{(N\!-\!2)(N\!-\!1)} \!\!&\!\! \zeta^{(N-1)^2}
\end{array} \!\! \right]\!\!\! \\[2pt]
\zeta &= e^{j2 \pi / N},
\end{align}
%
\noindent the IGFT is identical to the IDFT.

Thus, in the case of a ring graph, the GC is identical to the definition of the cepstrum.
Moreover, it is also identical to the definition of the spatial cepstrum of circular symmetric microphones in an isotropic sound field \cite{Imoto_TASLP2017_01}.
This means that the ring connection in the GC domain corresponds to the circular symmetric arrangement of microphones in an isotropic sound field in the acoustic spatial condition.
%
%
\section{Experiments}
\label{sec:Experiments}
\subsection{Experimental Conditions}
\label{ssec:cond}
\begin{figure}[t!]
\begin{center}
\includegraphics[width=0.72\columnwidth]{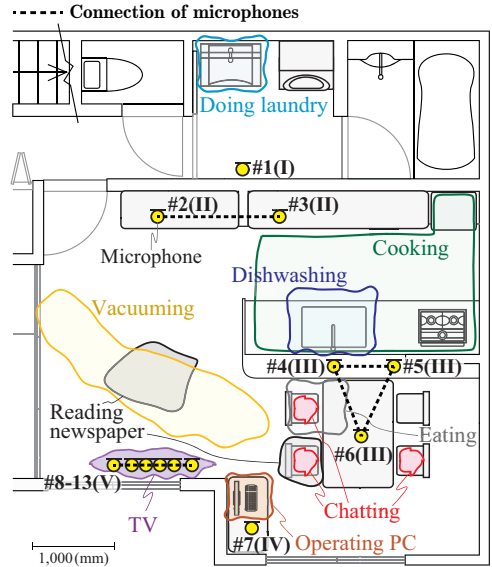}
\vspace{-7pt}
\end{center}
\caption{Microphone arrangement and sound source locations.
Channel indices (1--13) and group indices of synchronized microphones (I--V) are also indicated. }
\label{fig:condition11}
\end{figure}

To evaluate the effectiveness of the proposed method for partially synchronized microphones, we conducted classification experiments on acoustic scenes in a living room.
Since most of the public datasets for acoustic scene analysis including TUT Acoustic Scenes 2017 \cite{Mesaros_DCASE2017_01} and AudioSet \cite{Gemmeke_ICASSP2017_01} are provided in single or stereo channels, we recorded a multichannel sound dataset with 13 synchronized microphones in a real environment.
The sound dataset includes nine acoustic scenes, ``vacuuming,'' ``cooking,'' ``dishwashing,'' ``eating,'' ``reading a newspaper,'' ``operating a PC,'' ``chatting,'' ``watching TV,'' and ``doing the laundry,'' which happen frequently around the living room.
The microphone arrangement and the locations of the sound sources are shown in Fig.~\ref{fig:condition11}.
The recorded sounds consisted of 257.1 min. of recordings, which were randomly separated into 5,180 sound clips for model training and 2,532 sound clips for classification evaluation, where no acoustic scene overlapped with another scene in all the sound clips.
To evaluate the scene classification performance with synchronization mismatch among the microphone groups, the recorded sounds for classification evaluation were misaligned with various error times among the microphone groups shown in Fig.~\ref{fig:condition11}.
The error times were randomly sampled from a Gaussian distribution with $\mu = 0$ and various variances $\sigma^{2}$.
The other recording conditions and experimental conditions are listed in Table~\ref{tab:Condition}.
%
%
\begin{table}[t!]
\caption{Experimental conditions}
\label{tab:Condition}
\small
\centering
\renewcommand{\arraystretch}{0.98}
\begin{tabular}{lll}
\wcline{1-3}\\[-12pt]
\multicolumn{2}{l}{}&\vspace{-5pt} \\
\multicolumn{2}{l}{Sampling rate}&48 kHz\\
\multicolumn{2}{l}{Quantization bit rate}&16 bits\\
\multicolumn{2}{l}{Sound clip length}&8 s\\
\multicolumn{2}{l}{Frame length\hspace{1pt}/\hspace{1pt}FFT point}&20 ms\hspace{1pt}/\hspace{1pt}2,048\\
\multicolumn{2}{l}{Connection weight $\alpha$}&0.01\\
\cline{1-3}\\[-7pt]
\multicolumn{2}{l}{Network structure of CNN}&3 conv. \& 3 dense layers\\
\multicolumn{2}{l}{Pooling in CNN layers}&2 $\times$ 2 max pooling\\
\multicolumn{2}{l}{Activation function}&ReLU, softmax (output layer)\\
\multicolumn{2}{l}{\# channels of CNN}&32, 24, 16\\
\multicolumn{2}{l}{\# units of dense layers}&128, 64, 32\\
\multicolumn{2}{l}{Optimizer}&Adam\\
\wcline{1-3}
\end{tabular}
\end{table}
%
%
%
\subsection{Spatial Information Extracted by Graph Cepstrum}
\label{ssec:spatialinfo}
\begin{figure}[t!]
\begin{center}
\includegraphics[width=0.82\columnwidth]{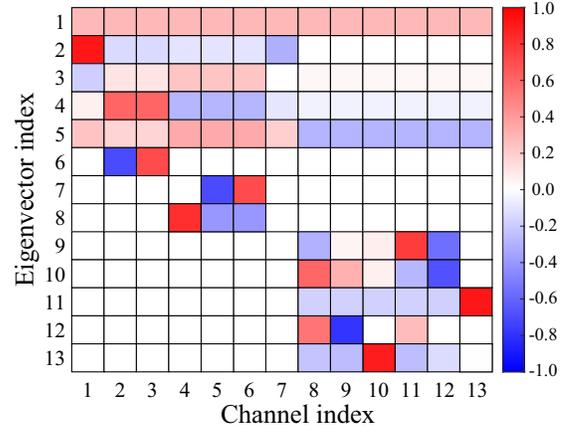}
\vspace{-11pt}
\end{center}
\caption{IGFT matrix ${\bf U}^{\mathsf{T}}$ in red-blue color map representation}
\label{fig:eigenvector}
\end{figure}

To clarify how the GC extracts spatial information, we show the IGFT matrix ${\bf U}^{\mathsf{T}}$ in Fig.~\ref{fig:eigenvector}.
The $n$th-row vector of ${\bf U}^{\mathsf{T}}$ corresponds to the $n$th-eigenvector of the graph Laplacian ${\bf L}$.
The $n$th-order GC is calculated using the $n$th-row vector of ${\bf U}^{\mathsf{T}}$ as follows:

\vspace{-15pt}
\begin{align}
e_{\tau, n} = {\bf u}_{n}^{\mathsf{T}} {\bf q}_{\tau} = \sum_{m=1}^{N} u_{n, m} q_{\tau, m},
\end{align}
\vspace{-6pt}

\noindent where $e_{\tau, n}$, ${\bf u}_{n}^{\mathsf{T}}$, $u_{n, m}$, and $q_{\tau, m}$ are the $n$th-order GC, the $n$th-row vector of ${\bf U}^{\mathsf{T}}$, the $(n, m)$ entry of ${\bf U}^{\mathsf{T}}$, and the $m$th element of ${\bf q}_{\tau}$, respectively.
This indicates that the $n$th-order GC is obtained by a linear combination of log-amplitudes $q_{\tau, m}$, where $u_{n,m}$ is the weight of the linear combination.
From Fig.~\ref{fig:eigenvector}, it can be interpreted that the first-order GC represents the average sound level in the whole space because all the weights ${\bf u}_{n}^{\mathsf{T}}$ are positive.
For the middle-order eigenvectors, the signs of the weights between connected microphones are similar.
This indicates that the GC can capture spatial information while taking the connections of microphones into account.
For the higher-order eigenvectors, the weights of only part of the connected microphone group are active and the signs of the weights differ.
These eigenvectors capture spatial information of the sound sources close to the microphone groups because if the sound sources are far from the microphone groups, the linear combination of the microphone groups is canceled in Eq. (13).
%
%
%
\subsection{Acoustic Scene Classification}
\label{ssec:classification}
Acoustic scenes were then modeled and classified for each sound clip using a Gaussian mixture model (GMM), a supervised acoustic topic model (sATM) \cite{Imoto_IEICE2016_01,Imoto_EUSIPCO2017_01}, and a convolutional neural network (CNN).
Specifically, the GMM was applied to acoustic feature vectors ${\bf e}_{\tau}$ and ${\bf d}_{\tau}$ for each acoustic scene $x$.
After that, acoustic scene $x$ of sound clip $c$ was estimated by calculating the product of the likelihoods over the sound clip as follows:

\vspace{-12pt}
\begin{align}
  x_{c}=\argmax_{x} \prod^{T_{c}}_{\tau=1} p_{\tau}({\bf f}_{\tau}|x),
  \label{eq:argmax}
\end{align}
\vspace{-7pt}

\noindent where $T_{c}$, ${\bf f}_{\tau}$, and $p_{\tau}({\bf f}_{\tau}|x)$ are the number of frames in sound clip $c$, an acoustic feature vector calculated frame by frame, such as ${\bf d}_{\tau}$ or ${\bf e}_{\tau}$, and the likelihood of acoustic scene $x$ at time frame $\tau$, respectively.
As other methods for acoustic scene classification utilizing a distributed microphone array, we also evaluated classifiers based on late fusion-based classification methods \cite{Kurby_Dcase2016_01}.
\begin{figure}[t!]
\begin{center}
\includegraphics[width=1.0\columnwidth]{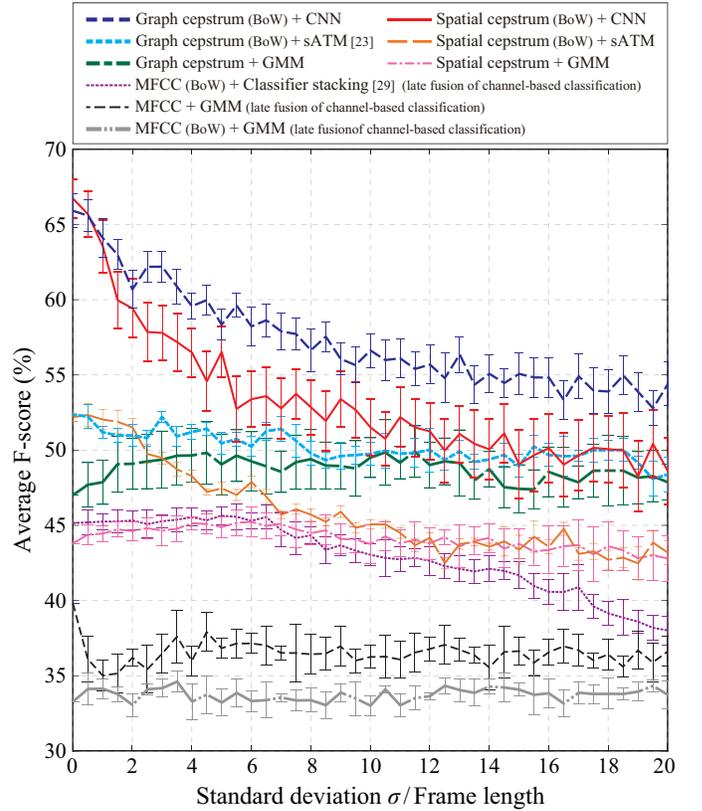}
\vspace{-12pt}
\end{center}
\caption{Acoustic scene classification accuracy with various synchronization error times between connected microphone groups}
\label{fig:gcres01}
\end{figure}
%
%
\subsection{Experimental Results}
\label{ssec:result}
The classification performance of acoustic scenes is shown in Fig.~\ref{fig:gcres01}.
For each experimental condition, the acoustic scene modeling and classification were conducted ten times with various synchronization error times sampled randomly.
These results show that when the synchronization error between microphone groups is small, the GC and conventional spatial cepstrum effectively classify acoustic scenes.
When the synchronization error between microphone groups increases, the scene classification performance for the GC slightly decreases.
In contrast, the classification accuracy decreases rapidly when using conventional methods.
This indicates that the proposed GC is more robust against synchronization error than conventional methods.
%
%
\section{Conclusion}
\label{sec:conclude}
%
In this paper, we proposed an effective spatial feature extraction method for acoustic scene analysis using partially synchronized or closely located distributed microphones.
In the proposed method, we derived the graph cepstrum (GC), which is defined as the inverse graph Fourier transform of the logarithm power of a multichannel observation.
We then demonstrated that the GC in a ring graph is identical to the conventional cepstrum and spatial cepstrum in a circularly symmetric microphone arrangement with an isotropic sound field.
Our experimental results using real environmental sounds showed that the GC more robustly classifies acoustic scenes than conventional spatial features even when the synchronization mismatch between partially synchronized microphone groups is large.
%
%
\section*{Acknowledgments}
Part of this work was supported by the Support Center for Advanced Telecommunications Technology Research, Foundation.
%
\vspace{-4pt}
\setlength\itemsep{-20pt}
\bibliographystyle{IEEEbib}
\bibliography{IEEEabrv,EUSIPCO2018ref,KeisukeImoto05}
%
\end{document}